\documentclass[aps,preprint,showpacs,superscriptaddress,groupedaddress,11pt,a4paper]{revtex4}

\usepackage{dcolumn}
\usepackage{bm}
\usepackage{natbib}
\usepackage{amsmath}
\usepackage{array}
\usepackage{xcolor}
\usepackage{hyperref}
\usepackage{neuralnetwork}
\usepackage{tikz}
\usepackage{mwe}
\usepackage{graphbox}
\usepackage{graphicx,pdflscape}
\usepackage{rotating}
\usetikzlibrary{matrix,chains,positioning,decorations.pathreplacing,arrows}
\usetikzlibrary{positioning}

\usepackage{xpatch}
\makeatletter
\xpatchcmd{\linklayers}{\nn@lastnode}{\lastnode}{}{}
\xpatchcmd{\linklayers}{\nn@thisnode}{\thisnode}{}{}
\makeatother

\begin{document}


\title{Learning and correcting non-Gaussian model errors}

\author{Danny Smyl}
 \altaffiliation{Department of Civil and Structural Engineering, University of Sheffield, UK}
 \email{d.smyl@sheffield.ac.uk}
 \author{Tyler N. Tallman}
 \altaffiliation{School of Aeronautics and Astronautics, Purdue University, West Lafayette, IN, USA}
  \email{ttallman@purdue.edu}
   \author{Jonathan A. Black}
   \email{j.a.black@sheffield.ac.uk}
 \altaffiliation{Department of Civil and Structural Engineering, University of Sheffield, UK}
\author{Andreas Hauptmann}
\altaffiliation{Research Unit of Mathematical Sciences, University of Oulu; Department of Computer Science, University College London}
\email{andreas.hauptmann@oulu.fi}
\author{Dong Liu}
\altaffiliation{CAS Key Laboratory of Microscale Magnetic Resonance and Department of Modern Physics, University of Science and Technology of China (USTC), Hefei 230026, China}
\altaffiliation[$^\diamond$]{Hefei National Laboratory for Physical Sciences at the Microscale, USTC, China}
\altaffiliation[$^\ddagger$]{Synergetic Innovation Center of Quantum Information and Quantum Physics, USTC, China}
 \email{dong2016@ustc.edu.cn}

\date{\today}
             
\raggedbottom 

\begin{abstract}
All discretized numerical models contain modelling errors -- this reality is amplified when reduced-order models are used.
The ability to accurately approximate modelling errors informs statistics on model confidence and improves quantitative results from frameworks using numerical models in prediction, tomography, and signal processing.
Further to this, the compensation of highly nonlinear and non-Gaussian modelling errors, arising in many ill-conditioned systems aiming to capture complex physics, is a historically difficult task.
In this work, we address this challenge by proposing a neural network approach capable of accurately approximating and compensating for such modelling errors in augmented direct and inverse problems.
The viability of the approach is demonstrated using simulated and experimental data arising from differing physical direct and inverse problems.
\end{abstract}

\pacs{PACS classification: 89}
\maketitle


\section{Introduction} Discretized numerical models, such as the spectral element and finite element method (FEM), are widely used to simulate physical systems that vary in space or space and time \cite{surana2016,komatitsch2002spectral}.
Likely the most familiar use of discretized numerical methods is in solving partial differential equations (PDEs).
The motivation for discretizing and solving PDEs numerically is broad but is largely centered on estimating solutions to problems with arbitrary geometry and boundary conditions that may otherwise be difficult to model analytically.
Numerical solutions to PDEs also enrich understanding on the spatial-temporal evolution of complex physics, for example in applications of black hole dynamics \cite{baker2006binary}, geophysical fluid dynamics \cite{natale2016compatible}, and elastoplasticity \cite{zervos2001finite}.

\begin{figure*}
\centering
\hspace{-17cm}\includegraphics[width=17cm]{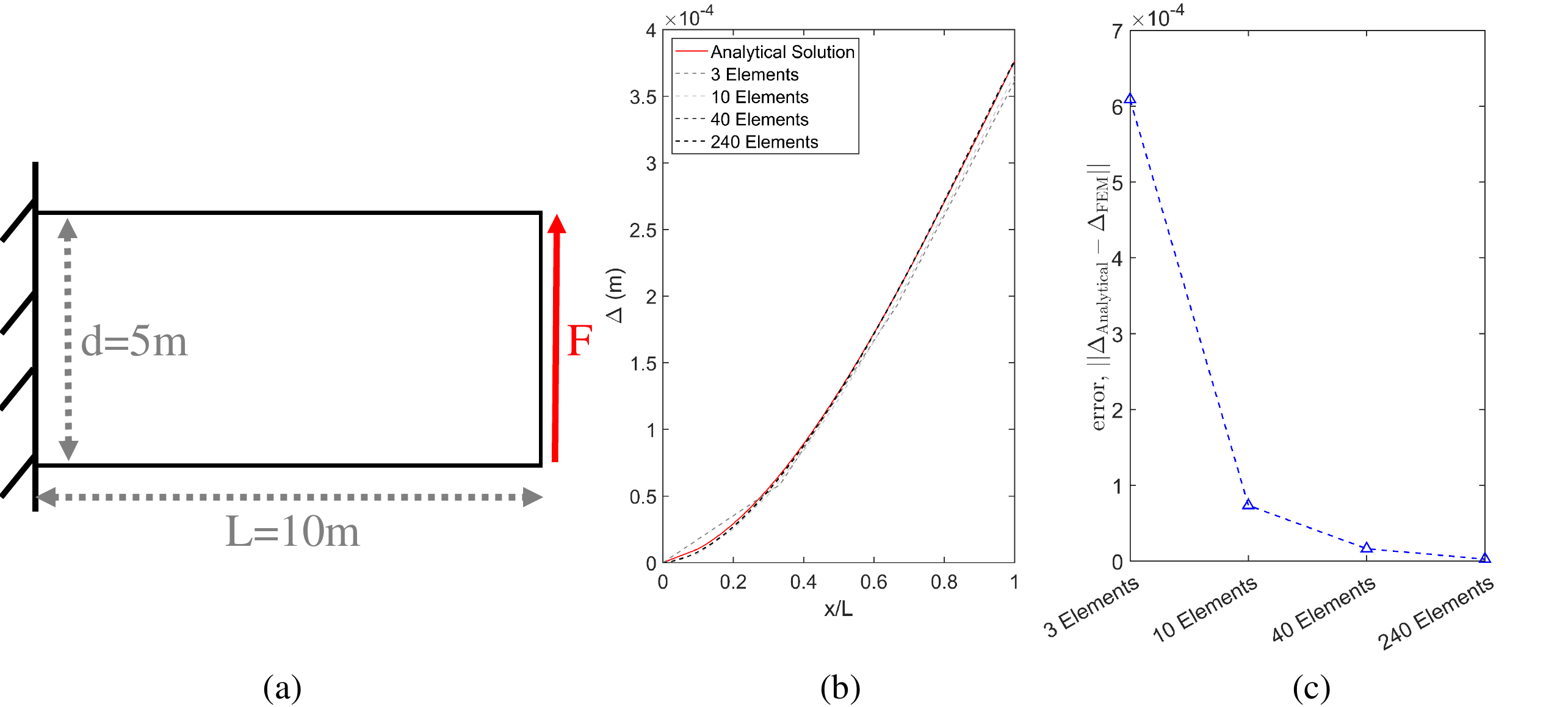}
\hfill
\caption{Illustrative example considering shear-deformable (Timoshenko) beam bending: (a) schematic illustration of the geometry considered, (b) analytical and $h$-refined FE displacement field solutions, and (c) $L_2$ norm errors of the FE displacement fields.}
\label{Exone}
\end{figure*}

Of course, this flexibility comes at a cost -- pervasive modelling errors.
These errors, largely a consequence of simplifying assumptions and the discretization refinement itself \cite{babovic2005error},  corrupt numerical solutions and are the source of frustration for many scientists and engineers.
Adding to the complexity of this situation, users often need to weigh the fidelity and resolution of numerical solutions against the expectation of errors (which are, themselves, uncertain).
To illustrate this phenomenon, consider a classical example where the FEM equipped with quadratic quadrilateral (Q8) elements is used to compute the displacement field of a fixed Timoshenko beam.
To investigate the modelling errors, we can choose varying levels of discretization refinement (i.e. $h$ refinement) and compare/benchmark to the analytical solution for the displacement field $\Delta$ as a function of the spatial coordinates -- horizontal and vertical coordinates $x$ and $y$, respectively. 
Using the analytical solution, the vertical displacements for a fictitious beam shown in Fig. \ref{Exone}(a) are given by $\Delta(x,y=0) = \frac{F}{6\,E\,I}\left[x^2\,\left(3\,L-x\right)+\frac{x\,\left(5\,\nu+4\right)d^2}{4}\right]$ \cite{augarde2008use}, where $F=2$MN is the end load, $L=10$m is the beam length, $E=200$MPa
is the modulus of elasticity, $\nu=0.33$ is the Poisson ratio, $I$ is the moment of inertia, and $d=5$m is the  depth of the beam's rectangular cross section. 
We show the trial solutions to this problem in Fig. \ref{Exone}(b), where the accuracy of the finite element (FE) solutions are indicated by their deviations from the analytical solution and further quantified in \ref{Exone}(c) using a standard $L_2$ error norm plot.
The illustrative problem here typifies the expected error convergence with increasing $h$ refinement -- i.e. as the mesh size densifies, the FE errors approach zero asymptotically.
This realization reinforces the fact that, because FE sizes cannot be infinitesimal, numerical modelling errors are unavoidable (although, for this simple example, error convergence to machine precision is possible).

Not surprisingly, significant research has been conducted to reduce model errors dating back to the inception of computing in science and engineering, especially by those working in inverse problems where modelling errors result in artifacts in medical, geophysical, and material imaging \cite{kaipio2007statistical}. 
Modelling errors are also a serious source of consideration for engineers using FE solutions to design infrastructure.
For example, in some engineering cases, 5 or 10 percent modelling error may be tolerable; however, engineering judgement and discretion often play key roles in such a design process.
Conversely, when the uncertainty of modelling error is sufficiently high or cannot be reliably estimated, minimizing errors may be the only option.
Appropriately, researchers in these fields have developed a number of regimes for understanding modelling errors due to, e.g., the well-known physically-unrealistic zero energy modes in low-order elements and mesh-size dependence \cite{zienkiewicz2006background}.

General approaches to approximating modelling errors -- ultimately leading to their subtraction from the discretized solution -- remain, however, very challenging \cite{smyl2019less}. 
Without question, authors such as Kaipio, Arridge, and coauthors have made substantial progress towards approximating assumed Gaussian modelling errors \cite{nissinen2010compensation,lehikoinen2007approximation,arridge2006approximation}, while benefiting from improved-accuracy model reduction.
Meanwhile, other modelling error approximation approaches have been effective in linear situations \cite{nievergelt1994total}.
In cases where modelling errors are not linear as a function of the PDE solution or the error statistics are unknown \emph{a priori}, available and accessible methods for approximating numerical modelling errors are sparse. First advances have been made for linear inverse problems, to incorporate an \emph{implicit} model correction \cite{hauptmann2018approximate} within a learned gradient descent scheme \cite{adler2017solving}. \emph{Explicit} corrections, as we investigate here, have been recently analyzed in \cite{lunz2020learned}, where the authors show that under sufficient approximation accuracy, we can expect to obtain solutions close to what we obtain with a correct model. We will take this as motivation and shall take the step to more demanding nonlinear inverse problems and verify the proposed techniques to experimental data. In particular, the approach discussed here provides an explicit quantification of the model error and not only a correction.

Considering first the realization that PDEs are themselves often highly nonlinear or that their solutions are nonlinearly dependent on the input parameters, one may quickly surmise that a ``one size fits all'' approach to quantifying modelling errors is possibly intractable.
However, delving more deeply into how PDEs are numerically solved, one observes that there are two broad schools of approaches: namely, physics-based models and physics-informed models.
The latter employs, e.g., neural networks (NNs) to develop surrogate models which emulate physics-based models and generate PDE solutions that are valid within the space of the physics-based training data.
Such approaches have been the source of significant research in the past few years \cite{iten2020discovering,raissi2019physics,chen2018neural,raissi2018hidden}; this interest largely stems from the speed and accuracy of simulating physics with well-trained NNs.
Accordingly, the success of NNs in developing nonlinear maps from causalities to PDE solutions leads to the ambition that NNs can also be used to develop effective nonlinear maps from PDE solutions to errors in PDE solutions.
Assuming the former can be actualized, the use of NNs developed via deep learning for approximating numerical modelling errors has the distinct advantage that error statistics are not required \emph{a priori}.
It is worth remarking that, from a cautionary standpoint, the accuracy of NN predictions is heavily dependent on the training data.
Accordingly, significant care should be taken when generating training data and assessing the reliability of NN predictions of modelling error.

\section{Approximating numerical modelling errors using Neural Networks}
The aim of deep learning, and in particular the use of neural networks, is to develop a nonlinear mapping $\mathcal{A}$ between two parameter spaces \cite{arridge2019networks}. 
In the context of this work, we aim to learn the functional relation between discretized numerical PDE solutions $u$ and model errors $\epsilon$ such that  $\epsilon = \mathcal{A}(u)$.
One advantage of using NNs is that there are numerous architectures and algorithms available for handling differing data structures characterized by $u$, for example convolutional networks in the case where $u$ has a spatial invariant structure but local neighborhood dependencies. 
Furthermore, there are a number of measures for $\epsilon$ which may be exploited to take advantage of specific network architecture properties.

Provided that we are interested in approximating modelling errors such that they may be directly compensated in numerical solutions, we define errors in the absolute sense.
This gives rise to errors $\epsilon = u_\mathrm{A} - u_\mathrm{R}$ where $u_\mathrm{A}$ is either accurate measured data or an accurate numerical model and $u_\mathrm{R}$ is a numerical model (when measurements are used) or a reduced-order numerical model.
Practically speaking, this error structure implies that the NN modelling error approximator would be used for the prediction 

\begin{equation}
\epsilon \approx \mathcal{A}(u_\mathrm{R})
\label{NNAE1}
\end{equation}

\noindent leading to a corrected model 

\begin{equation}
\widetilde{u}_\mathrm{A} \approx  u_\mathrm{R} + \mathcal{A}(u_\mathrm{R}).
\label{NNAE2}
\end{equation}

Further,  in cases where $u_\mathrm{A}$ is measured, $\epsilon$ is the true discrepancy between a physical numerical model and reality (less measurement noise).
Conversely, when $u_\mathrm{A}$ is a highly-accurate numerical model (e.g. a high-order $hp$ FEM discretization \cite{smyl2019less}) and $ u_\mathrm{R}$ is reduced order model (e.g. a low-order $hp$ FEM discretization), $\epsilon$ is an approximative discrepancy between the reduced order model and reality.
It is important to highlight both cases because, in many situations, high fidelity spatial or spatial-temporal physical measurements may not be available; therefore high-order models can be used as a substitute with certain application-specific limitations \cite{kaipio2006statistical}.
Moreover, the incorporation of a broader suite of physics can be readily included in high-order modelling that may not be pragmatically feasible in experimental environments.

\begin{figure*}[h!]
\centering
 \hspace{-15cm} \includegraphics[width=15cm]{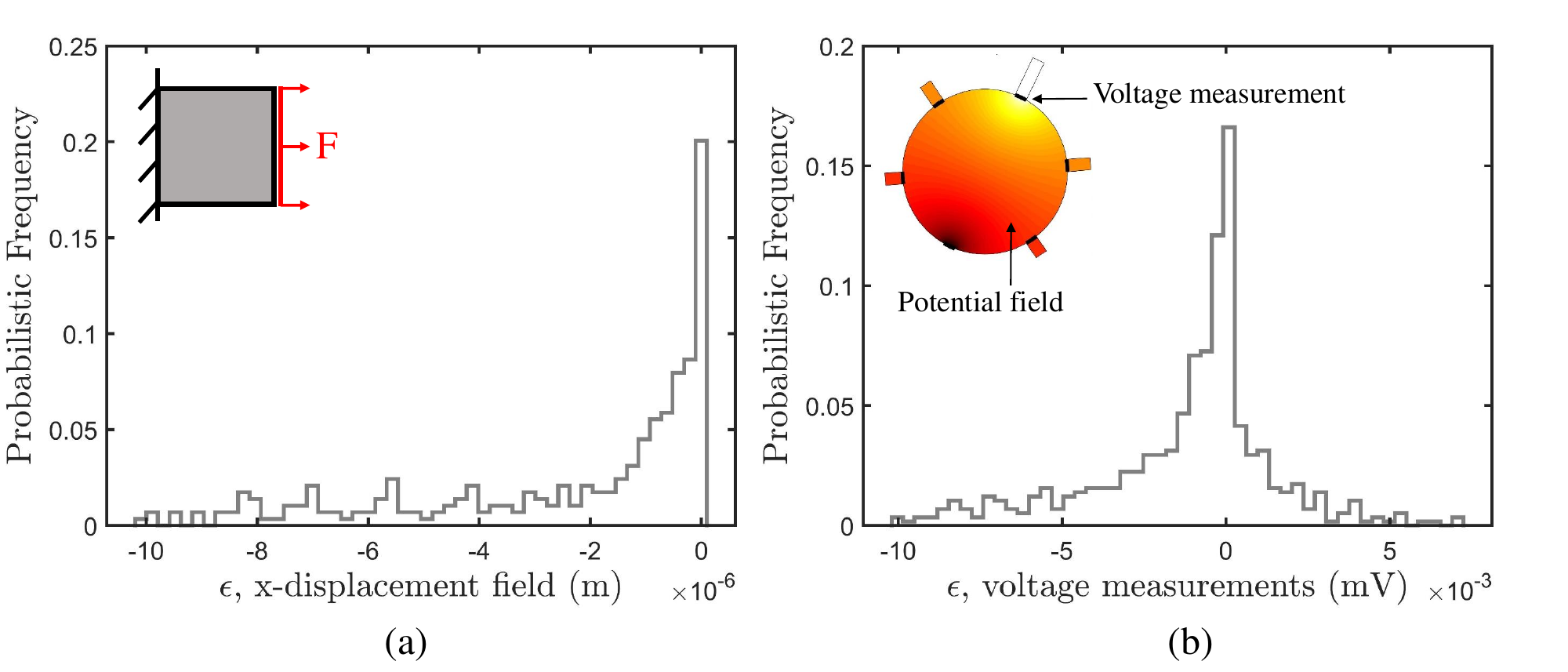} 
\hfill
\caption{Representative modelling error histograms showing the probabilistic error frequencies for (a) the nodal displacement field of a fixed-free square steel plate stretched in the horizontal ($x$) direction and (b) electrostatic boundary voltage measurements from an inhomogeneous conductivity circular domain. For case (a), $u_\mathrm{A}$ is computed using cubic square finite elements with 16 nodes and $u_\mathrm{R}$ is computed using 4-node square finite elements.  For case (b), $u_\mathrm{A}$ is computed using cubic triangular finite elements with 10 nodes and $u_\mathrm{R}$ is computed using 3-node linear triangular finite elements. In both cases, modelling errors are highly dissimilar and not well characterized by Gaussian distributions.}
\label{hist}
\end{figure*}

In developing this framework, there are no intrinsic assumptions with regards to the structure of the modelling errors.
Put plainly, we make no presupposition that $\epsilon$ has an inherent statistical distribution (Gaussian, Poisson, bi-modal, etc.).
This is a key point since modelling errors can have arbitrary distributions owing to, e.g., the non-homogeneity of material coefficients, boundary conditions, and non-linearity of the underlying physics.
To illustrate this, probabilistic modelling error histograms are shown for different numerical solutions in Figs. \ref{hist}(a) and \ref{hist}(b) exhibiting dissimilar and non-Gaussian distributions.
Finally, it is worth remarking that the nodal locations of points may vary between spaces $u_\mathrm{A}$ and $u_\mathrm{R}$.
When this occurs, we generally have to apply an interpolation function, inducing interpolation errors.
While these errors are not direct numerical modelling errors, they can be lumped into the proposed framework.
Nonetheless, the presence of interpolation errors further skews possible assumptions on what the statistical variation of numerical errors \emph{might} be -- placing further emphasis on the realization that making prior statistical assumptions on errors is not always favorable.

\section{Modelling error compensated inverse problems}  
A number of inverse problems, particularly those in geophysical and medical imaging, suffer from modelling errors.  
The presence of such error often results in imaging artifacts and degradation of spatial resolution.
Therefore, the inclusion of reliable modelling error approximators are of significant practical value.
Generally, solutions to inverse problems with a parameterization $\theta$ are solved by iteratively minimizing a cost functional of the form

\begin{equation}
\Psi = ||d - u_\mathrm{A}(\theta) ||^2 + R(\theta)
\label{cfa}
\end{equation}

\noindent where $d$ is a vector of experimental data, $R$ is a regularization functional to stabilize inversion and incorporates prior information, and $|| \cdot ||$ is the $L_2$ norm.
Realistically, highly-accurate models $u_\mathrm{A}$ are rarely (if ever) used owing to the computational expense of using them.
However, if we allow $\widetilde{u}_\mathrm{A} \approx u_\mathrm{A}$, we can substitute the model error correction terms as follows

\begin{equation}
\Psi  \approx ||d -  (u_\mathrm{R}(\theta) + \mathcal{A}(u_\mathrm{R}(\theta) ))||^2 + R(\theta).
\label{cfac}
\end{equation}

While a number of regimes can be used to solve the optimization problem in Eq. \ref{cfac}, we select the iterative Gauss-Newton method herein owing to its flexibility to handle a variety of prior models.
To do this, we update solutions $\theta_k$ at step $k$ using a linesearch to determine the step size in $\theta_k = \theta_{k-1} + s_k \Delta \theta$.
The minimizer $\Delta \theta$ includes information on the derivative of the numerical model, in this case $\widetilde{u}_\mathrm{A}$.
This information is contained in the error-compensated Jacobian $J$ and can be written in the iterate form 

\begin{equation}
J = \frac{\partial \widetilde{u}_\mathrm{A}(\theta_k)}{\partial \theta_k} = 
\frac{\partial (u_\mathrm{R}(\theta_k) + \mathcal{A}(u_\mathrm{R}(\theta_k)))}{\partial \theta_k}
\label{J0}
\end{equation}

\noindent and be computed numerically using a number of perturbation-based methods.
Ultimately, this information is included in the minimizer and regularized with a matrix $L$ and hyperparameter $\alpha$ as 

\begin{equation}
\Delta \theta = (J^T J + \alpha L^T L )^{-1} J^T(d - \widetilde{u}_\mathrm{A}(\theta_k))
\label{min1}
\end{equation}

\noindent and in the error-compensated form,

\begin{equation}
\Delta \theta = (J^T J + \alpha L^T L )^{-1} J^T(d - (u_\mathrm{R}(\theta_k) + \mathcal{A}(u_\mathrm{R}(\theta_k)))).
\label{min1}
\end{equation}

\noindent A number of regularization matrices $L$ are available (cf. \cite{reichel2009simple} for possible choices) and are usually selected to include prior information related to the problem physics.
For example, solutions may be assumed smoothly spatially distributed or sparse and would therefore change the selection of $L$.
Lastly, the selection of $\alpha$ has significant implications on solutions.
In the $L_2$ sense used herein, $R(\theta) = \alpha||L \theta ||^2$ implies that $\alpha$ controls the weighting between the data fidelity and regularization terms.
As such, we use L-curve analysis in selecting $\alpha$ in order to rationally optimize the weighting between data fidelity and regularization terms.

It is important to remark here that the proposed error-compensated inverse problem framework only accounts for errors in the reduced order forward model and its Jacobian.
This is in contrast to recent works, such as \cite{nissinen2010compensation,lehikoinen2007approximation,arridge2006approximation}, where researchers have used projection operators to ensure solutions lie in the desired space.
Commonly, the use of error covariance matrices (and their Cholesky factors) are used for this purpose.
In this work, however, we consider only the direct compensation of numerical modelling errors in reduced order models.


\section{Network structuring}  The purpose of tailoring a NN to a specific application, in the context of this work, is to ensure the network is sufficiently robust to (a) make reliable predictions using validation data independent from the training data and (b) also ensure the NN is able to make predictions within the space of the training data.
In this sense, training data can be considered as prior information on the anticipated space of NN predictions.
In this work, this is the space of modelling errors to be predicted encompassed by the suite of PDE boundary conditions and input parameters to be considered.

We consider fully connected, three hidden-layer feedforward networks for this initial research to satisfy requirements (a) and (b) for the problems studied herein.
In the network, we employ ReLU activation functions for the hidden layers,
linear activation functions for the output layer, and a $L_2$ regularization scheme such that the objective function is weighted as follows

\begin{equation}
\Phi = \frac{1}{N}\sum_{l=1}^N (\epsilon^d_l - \epsilon^\mathcal{A}_l)^2 + \lambda \Gamma
\label{NNr}
\end{equation}

\noindent where $ \Gamma=||w||^2$, $w$ are the network weights, $N$ is the number of samples, $\epsilon^d_l $ is the desired modelling error output and $\epsilon^\mathcal{A}_l$ is the output from the network at sample $l$.
As seen in the NN cost functional, Eq. \ref{NNr}, the selection of the hyperparameter $\lambda$ has an effect on the degree of data fitting and therefore biases the network.
This is dealt with by adopting a Levenberg–Marquard optimization approach, adaptively computing an optimal $\lambda$, to minimize Eq. \ref{NNr}.
In addition to the regularization already provided, a dropout of 25\% is included to improve the network robustness and generality.
Lastly, the implementation of the network is handled using Keras \cite{chollet2015keras}.

\begin{figure*}
\begin{neuralnetwork}[height=4.9]
    \newcommand{\nodetextclear}[2]{}
    \newcommand{\nodetextx}[2]{\ifnum #2=4 $u_\mathrm{R}$ \else $u_\mathrm{R}$ \fi}
    \newcommand{\nodetexty}[2]{\ifnum #2=4 $\epsilon$ \else $\epsilon$ \fi}
    \inputlayer[count=4, bias=false,exclude={3}, title=Input\\layer, text=\nodetextx]
    \hiddenlayer[count=5, bias=false, exclude={4}, title=Hidden\\layer]
     \linklayers[not to={4},not from={3}]
    \hiddenlayer[count=5, bias=false, exclude={4}, title=Hidden\\layer]
     \linklayers[not to={4},not from={4}]
         \hiddenlayer[count=5, bias=false, exclude={4}, title=Hidden\\layer]
     \linklayers[not to={4},not from={4}]
         \outputlayer[count=4, exclude={3}, title=Output\\layer, text=\nodetexty] 
      \linklayers[not to={3},not from={4}]

\path (L0-2) -- node{$\vdots$} (L0-4);		
\path (L1-3) -- node{$\vdots$} (L1-5);	
\path (L2-3) -- node{$\vdots$} (L2-5);	
\path (L3-3) -- node{$\vdots$} (L3-5);	
\path (L4-2) -- node{$\vdots$} (L4-4);	
\end{neuralnetwork}
\\
{\large(a)}
\\
\hspace{-15cm}
\includegraphics[width=15cm]{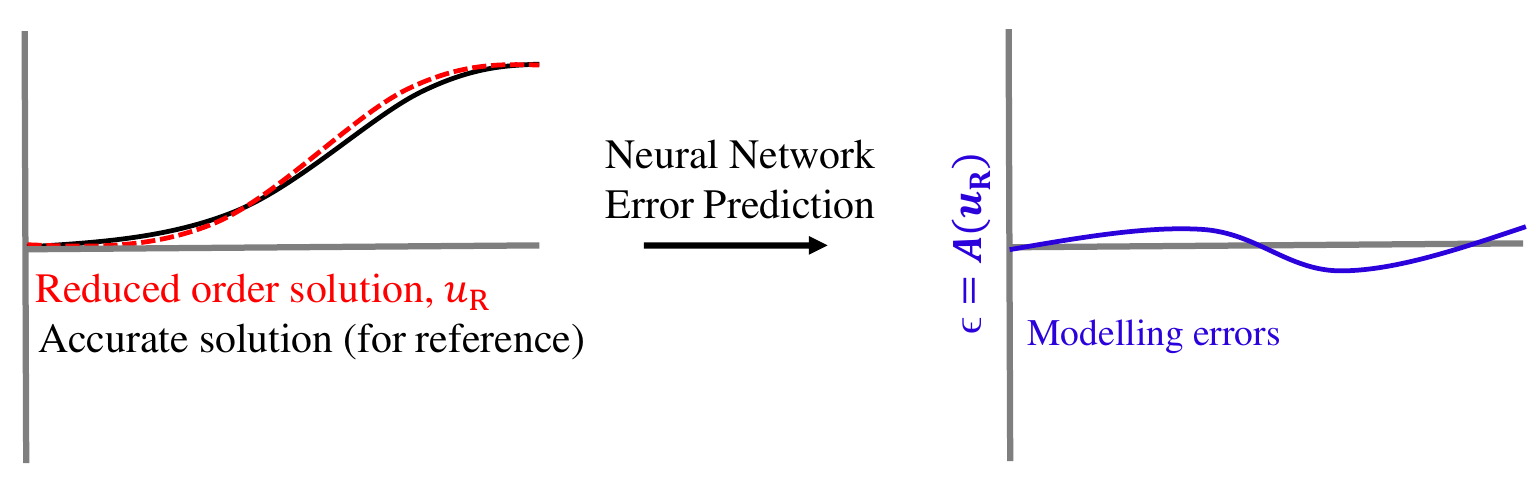}
\\
{\large(b)}
\hfill
\caption{Schematic illustrations of (a) the feedforward network used for the nonlinear mapping $\epsilon = \mathcal{A}(u_\mathrm{R})$ and (b) the trained neural network inputs and outputs. }
\label{nn1}
\end{figure*}
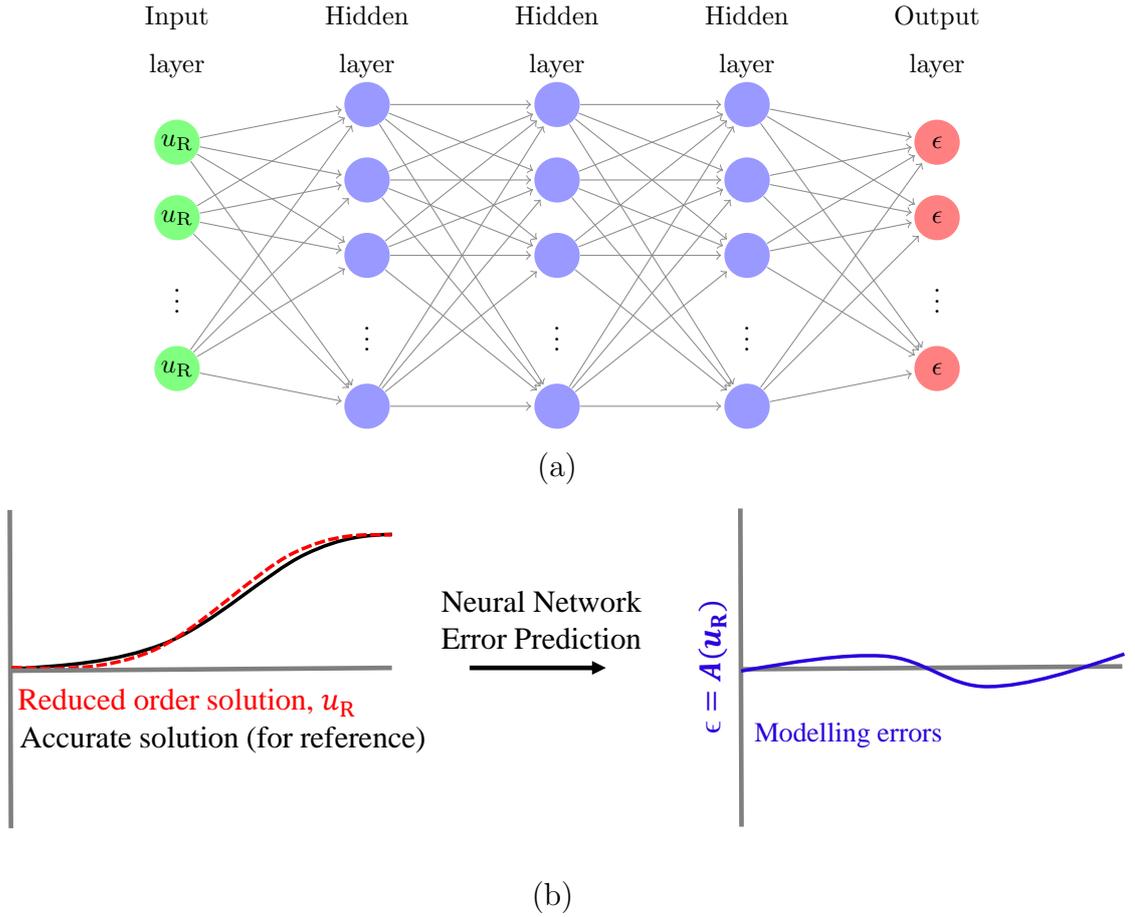

\section{Application of NN-Error Correction to Direct Problems -- Linear Elasticity}
\label{NMDEl}
Numerical problems in elasticity are pervasive in science and engineering; for example, in biophysics \cite{kalyanam2009poro} and multidisciplinary mechanical engineering \cite{askes2011gradient}.
Herein, we investigate modelling errors in the well-known two-dimensional linear elastic problem in a geometry $\Omega$ having a boundary $\partial \Omega$.
We define boundary tractions $\hat{f}$ on $\partial \Omega_{1}$, as part of the boundary.
Following, the displacements on $\partial \Omega_{2}$ are defined by $\hat{u}$.
Neglecting body forces, the elastic problem is written

\begin{equation}
\frac{\partial \sigma}{\partial x} = 0, \textbf{x} \in \Omega
\label{QSEI1}
\end{equation}

\begin{equation}
f(\textbf{x})  \equiv \sigma \bar{n}  \equiv \hat{f}, \textbf{x} \in \partial \Omega_{1}
\label{QSEI2}
\end{equation}

\begin{equation}
u = \hat{u}, \in \partial \Omega_{2}
\label{QSEI3}
\end{equation}

\noindent where $\sigma$ is the stress tensor, $\textbf{x}$ is the spatial coordinate, $\bar{n}$ is the unit normal, and $\partial \Omega_{1} \cup \partial \Omega_{1} \equiv \partial \Omega$.
To obtain a discretized FEM form, we adopt the weak Galerkin formulation of Eq. \ref{QSEI1} \cite{surana2016}.
From the discretized finite problem, we obtain a system of equations as a function of the elastic modulus $E$ given by $\sum_{i=1}^n K_{ij}({E})u_j=F_i$ where $K_{ij}$ is the FE stiffness matrix and $F_i$ is the applied force vector.
Because we are interested in computing the displacement field, this is inverted as $u_j = \sum_{i=1}^n K_{ij}^{-1}({E})F_i $ where $n$ is the number of unknown displacements.
Importantly, the accuracy of the FEM solution  is dependent on the element sizes $h$ and the degree of interpolating polynomials $p$ -- this is commonly referred to as the $hp$ FEM \cite{smyl2019less}, which is known to have exponential convergence of solution accuracy as $h$ is refined and $p$ is increased simultaneously.

Using the framework above, we can flexibly select the $h$ and $p$ spaces for FEM models and directly compare their solutions to compute modelling errors.
This is done by selecting $u_\mathrm{A}$ to be higher order than $u_\mathrm{R}$.
We investigate a fixed-free plate stretching problem where $\Omega$ is a square domain with dimensions 10m $\times$ 10m, a uniform thickness 0.01m, homogeneous Poisson ratio $\nu = 0.33$, and a uniformly-distributed outward boundary force on the free end $F=2$kN/m is applied. 

\begin{figure*}[h!]
\centering
 \hspace{-12.5cm} \includegraphics[width=12.5cm]{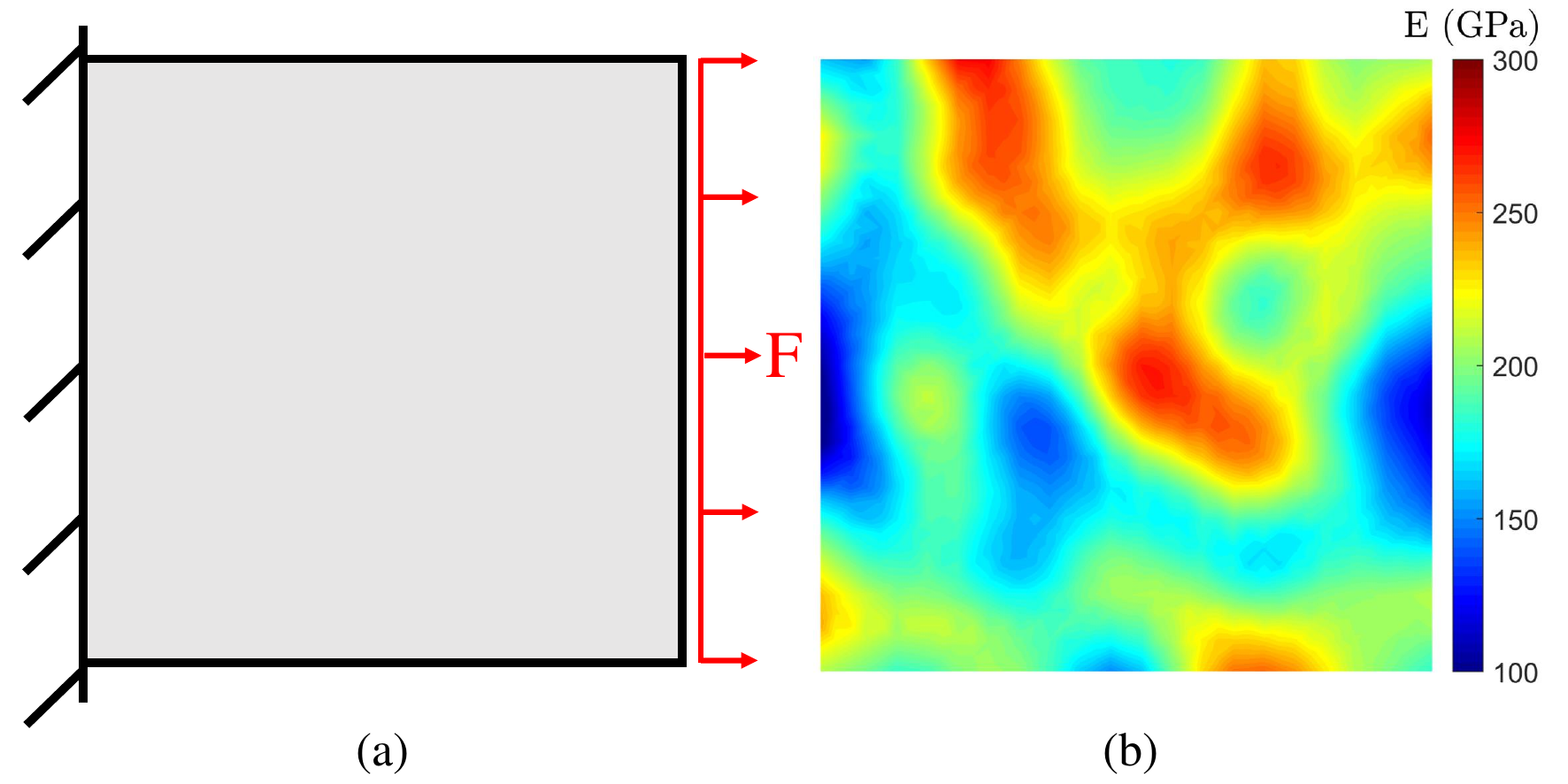} 
\hfill
\caption{Images showing (a) the problem setup and (b) a randomized inhomogeneous elastic modulus sample used in training for the direct linear elasticity problem.}
\label{ElasticSquare}
\end{figure*}

To generate training and validation data for the error approximation network, 20000 training and 5000 validation data series are computed.
Each data set includes displacement data $u_\mathrm{A}$ computed using 400 cubic structured elements with 16 nodes, $u_\mathrm{R}$ computed using 400 quadratic structured elements with 8 nodes, and absolute error data which is the difference between each data set.
In order to induce additional modelling errors, the elastic modulus of the plate is varied using random blobby distributions varying between 100 and 300GPa with a correlation length of 3m in each sample.
The range of the randomized elastic modulus used here has a mean value corresponding to that of steel and was selected based on its prevalence in numerical structural modelling.
For contextualization, a random sample is shown in Fig. \ref{ElasticSquare}(a).

\begin{figure}
\centering
\hspace{-9cm}
\includegraphics[width=9cm]{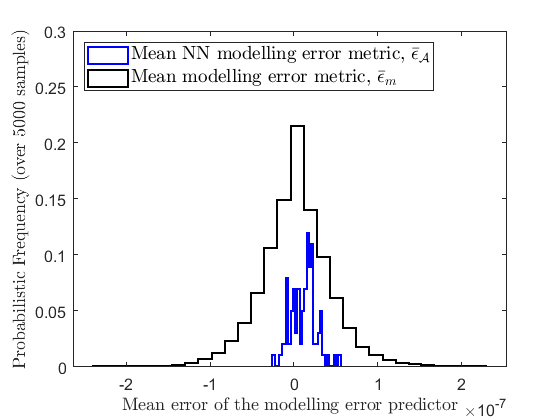}
\caption{Numerical experiment mean errors of the modelling error predictors, $\bar{\epsilon}_\mathcal{A}$ and $\bar{\epsilon}_\mathrm{m}$, which are computed from the NN and mean modelling error approximation methods.
$x$-axis values refer to the averaged deviations in modelling error predictions with respect to the true modelling errors calculated over 5000 validation samples.}
\label{elastic}
\end{figure}

After training the NN using 500 neurons per hidden layer, the NN error predictions are validated against the $v = 5000$ samples independent from the training data.
To do this, we compute the true errors from each validation data set $q \in v$ 
against the NN model error predictions.
This is done by subtracting these quantities for each $q$; we then obtain a metric for the average performance of the NN model error predictor $\bar{\epsilon}_\mathcal{A}=\frac{1}{v}\sum_q^v \epsilon_q - \mathcal{A}(u_{\mathrm{R},q})$.
For comparison, the mean modelling errors $\mu_\mathrm{m}=\frac{1}{v}\sum_q^v u_{\mathrm{A},q}-u_{\mathrm{R},q}$, commonly used in model error approximation, were also computed.
The mean modelling errors are then similarly subtracted from the true modelling errors to derive the metric $\bar{\epsilon}_\mathrm{m} = \frac{1}{v}\sum_q^v \epsilon_q - \mu_\mathrm{m}$.
Plainly speaking, the metrics $\bar{\epsilon}_\mathcal{A}$ and $\bar{\epsilon}_\mathrm{m}$ represent the ``mean errors of the error approximations" for the NN and the mean modelling error approaches, respectively.
A comparison of these metrics for the validation series is shown in Fig. \ref{elastic}.

Probabilistic frequency histograms shown in Fig. \ref{elastic} demonstrate the effectiveness of the NN modelling error approximation method.
Compared with the conventional mean model error approximation method, the errors in the NN predictions are much more tightly grouped around zero and have approximately the same frequency close to zero.
This observation means that, for the elastic problem considered, the NN approach to approximating modelling errors is more reliable than the conventional method since the deviation from zero is significantly smaller.
Regarding the closeness of the NN and mean modelling error approaches very near zero; this is effectively an artifact of the zero-displacement boundary conditions at the plate fixity.
As a consequence, the most sensible metric for evaluating Fig. \ref{elastic} is to simply consider the variance from the mean of each distribution -- visually, the NN approach clearly deviates less.
Broadly, the relative improvement of the NN approach is centered on the realization that modelling errors for the FE elasticity problem are not Gaussian (cf. Fig \ref{hist}(a)), and therefore the mean modelling error approach does not generalize as well as the NN predictor.
Preliminarily, this realization supports the potential use of NNs in predicting non-Gaussian modelling errors in discretized numerical models.

\begin{figure*}[h!]
\centering
 \hspace{-14cm} \includegraphics[width=14cm]{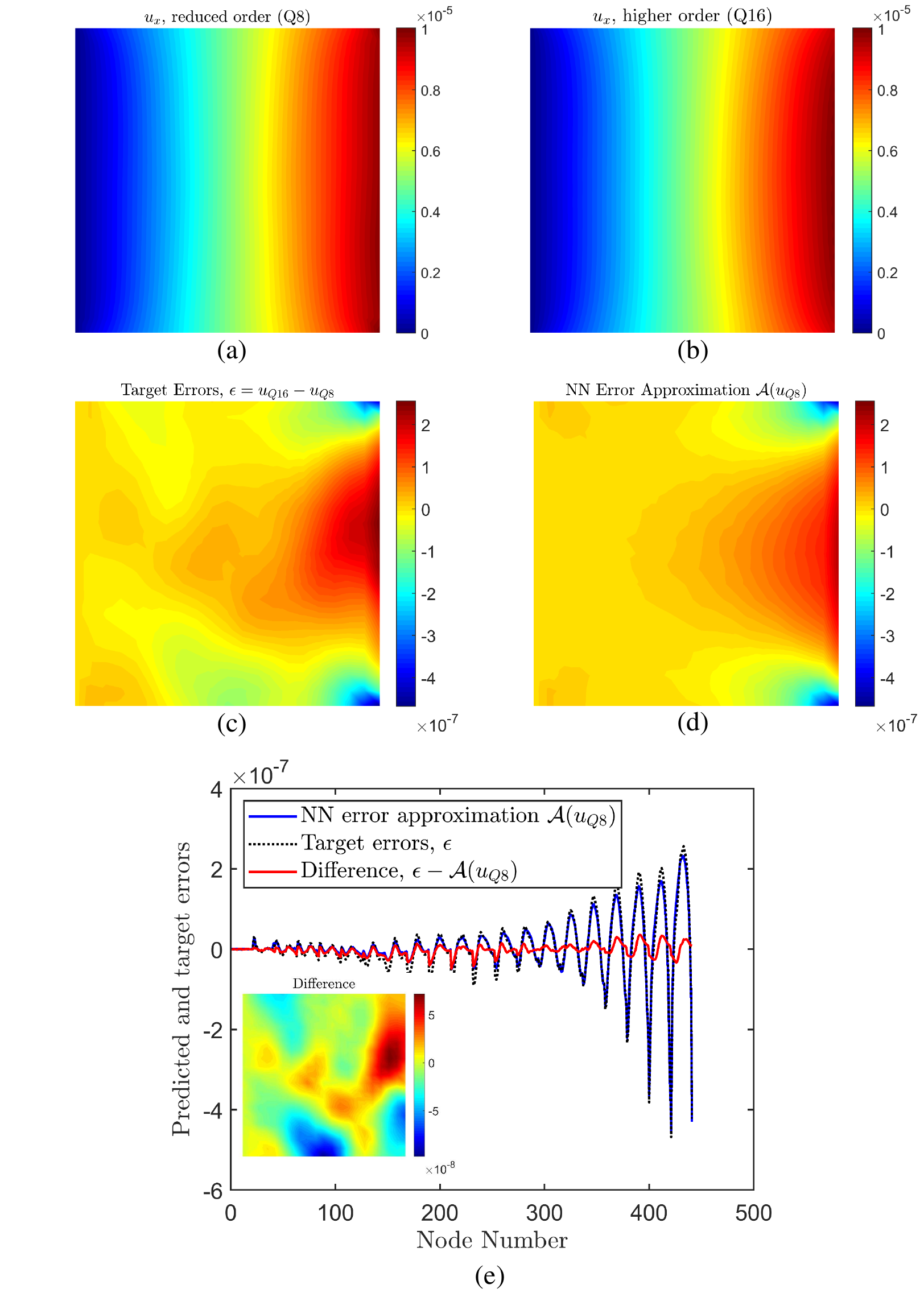} 
\hfill
\caption{Direct elasticity results showing (a) reduced order Q8 FE solutions $u_{Q8}$ for the $x-$displacement field, (b) higher order Q16 FE solutions $u_{Q16}$ for the $x-$displacement field, (c) target errors, (d) NN approximation [prediction] of the target errors, and (e) a line plot of the target errors, NN error approximation of the errors, and their difference evaluated the at the element corners.
}
\label{Elasticerri}
\end{figure*}

Having investigated the direct elasticity problem in a general sense, we now more closely examine a representative case sampled from the validation data.
To do this, we first show the $x$-displacement fields in Fig. \ref{Elasticerri}(a) and Fig. (b), which are mostly indistinguishable to the eye.
By taking their difference to obtain the target error distribution (as shown in Fig. \ref{elastic}(c)), we can better observe the differences between the reduced order Q8 and higher order Q16 finite element solutions.
We can then directly compare the target error distribution to the NN error prediction shown in Fig. \ref{Elasticerri}(d).
In doing this, we note that the qualitative spatial structure of the errors is reasonably well captured by the NN predictor near the plate's free end but deviates towards the fixity.
This result makes intuitive sense, since the magnitude of the displacements -- and their errors -- approach zero at the fixity. Therefore, errors near the fixity carry less weight during the NN training process and are consequently not as well predicted as errors near the free end.

To quantitatively analyze the former visual observations, the discrepancy between the target errors and the NN predicted errors is shown via line plots and a heat map in Fig. \ref{Elasticerri}(e).
Note that, for the purpose of clear visualization in Fig. \ref{Elasticerri}(e), only the element corner displacements are shown on the line plot.
As a whole, the previous visual observations are confirmed by the line plot.
Namely, since fluctuations in the difference between the target and predicted errors are roughly constant, the relative errors in larger displacements towards the free end are lower than near the fixity.
This realization highlights a central weakness of the proposed NN error approximation approach: that the relative accuracy of a predicted error at a given node is proportional to the magnitude of the reduced-order solution at that node.
Visually, this is observed in the heat map in Fig. \ref{Elasticerri}(e), where the noted differences fluctuate across the entire plate geometry.

\break

\section{Application of NN-Error Correction to Inverse Problems -- EIT} 
In this section we test the efficacy of the NN model error approximation approach in the context of electrical impedance tomography (EIT).
In EIT, the general aim is to reconstruct electrical conductivity $\gamma$ from distributed voltage measurements $V$.
Specifically, we consider a so-called self-sensing material with deformation-dependent electrical conductivity (i.e. the material is \emph{piezoresistive}) subject to an applied stress. The goal of EIT, therefore, is to recover the change in $\gamma$ due to this loading. This application is selected because (a) this problem is of novel importance in the non-destructive testing community, (b) the problem is notoriously difficult to solve using quantitative reconstruction approaches, and (c) this provides important experimental validation of the error correction herein proposed in the context of inverse problems.
The reasoning for (b) is owed to low sensitivity of $\gamma$ to deformation and the highly inhomogeneous background conductivity; thus, the presence of modelling  errors has a massively corrupting effect on reconstructions.
As a result, this problem is generally solved using subtractive difference imaging frameworks, where modelling errors are largely canceled via subtraction \cite{hassan2019failure}.

\subsection{EIT forward and inverse model description}
Before providing details on the inversion approach, we will first describe the EIT forward problem.
In EIT, the relation between the the electrical conductivity distribution $\gamma$ and numerically-modeled electrode voltages $U$ (i.e. the forward solution) is 
given by the complete electrode model (CEM)
which is formed by the partial differential equation

\begin{equation}
\nabla \cdot (\gamma \nabla u) = 0, \ \ \ x\in\Omega
\label{Eq.Poisson}
\end{equation}

 \noindent and the boundary conditions


\begin{equation}
u + \xi_l \gamma \frac{d u}{d\bar{n}} = U_l, \ \ \ 
x \in e_{\ell},\ \ell=1,\ldots,L
\label{bc1}
\end{equation}

\begin{equation}
\gamma \frac{d u}{d\bar{n}} =0, \ \ \ 
x\in \partial\Omega \backslash \bigcup_{\ell=1}^{L} e_{\ell}  
\label{bc2}
\end{equation}
\begin{equation}
\int_{e_l}\gamma\frac{d u}{d\bar{n}}\,dS\ = I_l, \ \ \ 
\ell=1,\ldots,L
\label{bc3}
\end{equation}

\noindent where 
$\Omega$ is the domain of the target, $\partial\Omega$ is the domain's boundary, $u$ is the electric potential, and
${e_l}$ is the ${l^{th}}$ electrode
\cite{somersalo92}.
Further,
${\xi_l}$, ${U_l}$ and ${I_l} $, respectively, are the contact impedance, electric potential, and electrical current corresponding to the ${l^{th}}$ electrode.
In addition to these, the current conservation law must be satisfied by writing 

\begin{equation}
\sum_{l=1}^L I_l = 0
\label{ccl}
\end{equation}

\noindent and the potential reference level is fixed, for example using

\begin{equation}
\sum_{l=1}^L U_l = 0.
\label{fp}
\end{equation}

\noindent In our formulation of the FEM framework, the FEM solution is obtained using an element-wise discretization of $\gamma$ with the $p-$adaptive EIDORS open source software \cite{adler2006uses}.

To solve the inverse problem, we utilize the so-called nonlinear difference imaging (NLDI) framework \cite{liu20155}, which is a specific case of the minimization problem described by Eq. \ref{cfa}.
Namely, the NLDI framework aims to reconstruct $\gamma_1$ (before a change in state) and the change in $\Delta \gamma$ (due to the change in state) by writing the following relation

\begin{equation}
 \gamma_2=\gamma_1 + \Delta \gamma
\label{newmodel}
\end{equation}

\noindent implying that the second conductivity state $\gamma_2$ results from an additive change in state.
Using this relation, the observation models for states before and after state change (at a state of zero stress and in a stressed state, respectively)
can be written as follows

\begin{equation}
\begin{matrix}
\begin{aligned}
V_1
&= U(\gamma_1) + \varepsilon_1 \\
V_2 
&= U(\gamma_1 + \Delta \gamma) + \varepsilon_2
\end{aligned}
\end{matrix}
\label{obmodel2}
\end{equation}

\noindent where $V_1$ and $V_2$ are the measurements before and after state change and $\varepsilon_1$ and $\varepsilon_2$ are the measurement noise.
By concatenating the observation models, we obtain a stacked observation model by writing

\begin{equation}
\underbrace{\begin{bmatrix}
V_1\\
V_2\\
        \end{bmatrix}}_\text{$\overline{V}$}
=
\underbrace{\begin{bmatrix}
         U(\gamma_1) \\
         U(\gamma_1 + \Delta \gamma)\\
        \end{bmatrix}}_\text{$\overline{U}(\bar{\gamma})$}
+
\underbrace{\begin{bmatrix}
         \varepsilon_1 \\
         \varepsilon_2\\
        \end{bmatrix}}_\text{$\overline{\varepsilon}$}
\label{newmodel1}
\end{equation}

\noindent which may be more written in a more compact form using

\begin{equation}
\overline{V} = \overline{U}(\bar{\gamma}) + \bar{\varepsilon}
\label{newmodel2}
\end{equation}

\noindent where 

\begin{equation}
\bar{\gamma}
= 
\begin{bmatrix}
         \gamma_1 \\
         \Delta \gamma\\
        \end{bmatrix}.
\label{newmodel3}
\end{equation}

Having now formalized the components required for inversion, we may directly substitute into Eq. \ref{cfa} to obtain the functional to be minimized in the NLDI approach (without NN error approximation)

\begin{equation}
\Psi_\mathrm{NLDI}(\gamma_1 > 0.3\gamma_\mathrm{ref}, \gamma_2 > 0) = ||\overline{V}  - \overline{U}(\bar{\gamma}) ||^2 + \bar{R}(\bar{\gamma})
\label{NLDIm}
\end{equation}

\noindent where $\bar{R}$ is a compound regularization functional and the constraints on the left hand side result from the realization that conductivity is a non-negative quantity and experimental testing.
For the latter, the constraint $\gamma_1 > 0.3\gamma_\mathrm{ref}$ is enforced to ensure the variance of the background conductivity is within a physically-realistic tolerance of approximately two standard deviations of the measured value $\gamma_\mathrm{ref}$.
Further, the EIT NLDI problem is well-known to be a nonlinear ill-posed problem and therefore requires regularization to ensure a conditionally well-posed solution and to stabilize inversion.
For this, the Laplacian operator is used for both states, resulting in $\bar{R} = \mathrm{blkdiag}[\alpha \nabla^2(\gamma_1),\alpha \nabla^2(\Delta \gamma)]$.
Similarly, the Jacobian also takes a concatenated from using 

\begin{equation}
J_\mathrm{NLDI} 
= 
\begin{bmatrix}
         J_{1} & \textbf{0} \\
         J_{2} & J_{2}\\
        \end{bmatrix}
        \label{Jstacked}
\end{equation}

\noindent where $J_{1}$ and $J_{2}$ are the Jacobians evaluated for $\gamma_1$ and $\gamma_2$.
Lastly, in order to modify Eq. \ref{NLDIm} to include model error approximation, we may simply substitute the EIT forward operator into the NN error approximator by letting $u_\mathrm{R} = \overline{U}(\bar{\gamma})$ and writing

\begin{equation}
\Psi_\mathrm{NLDI}^\mathcal{A}(\gamma_1 > 0.3\gamma_\mathrm{ref}, \gamma_2 > 0)  = ||\overline{V} -  
(\overline{U}(\bar{\gamma})
+ \mathcal{A}(\overline{U}(\bar{\gamma})))||^2 
+ \bar{R}(\bar{\gamma})
\label{NLDIae}
\end{equation}

\noindent where the superscript $\mathcal{A}$ on $\Psi_\mathrm{NLDI}$ denotes that the functional is modelling-error compensated.
Moreover, the Jacobian matrix (Eq. \ref{Jstacked}) for the modelling-error compensated model requires adjustment via Eq. \ref{J0}.
In the following subsection, use of the NN error compensated and non-error compensated NLDI frameworks will be employed in reconstructing conductivity changes in nanocomposite specimens subjected to applied stresses.

\subsection{NN error compensated reconstruction of conductivity changes in nanocomposites subjected to external stress}
\begin{figure}
\centering
\hspace{-9cm}
\includegraphics[width=9cm]{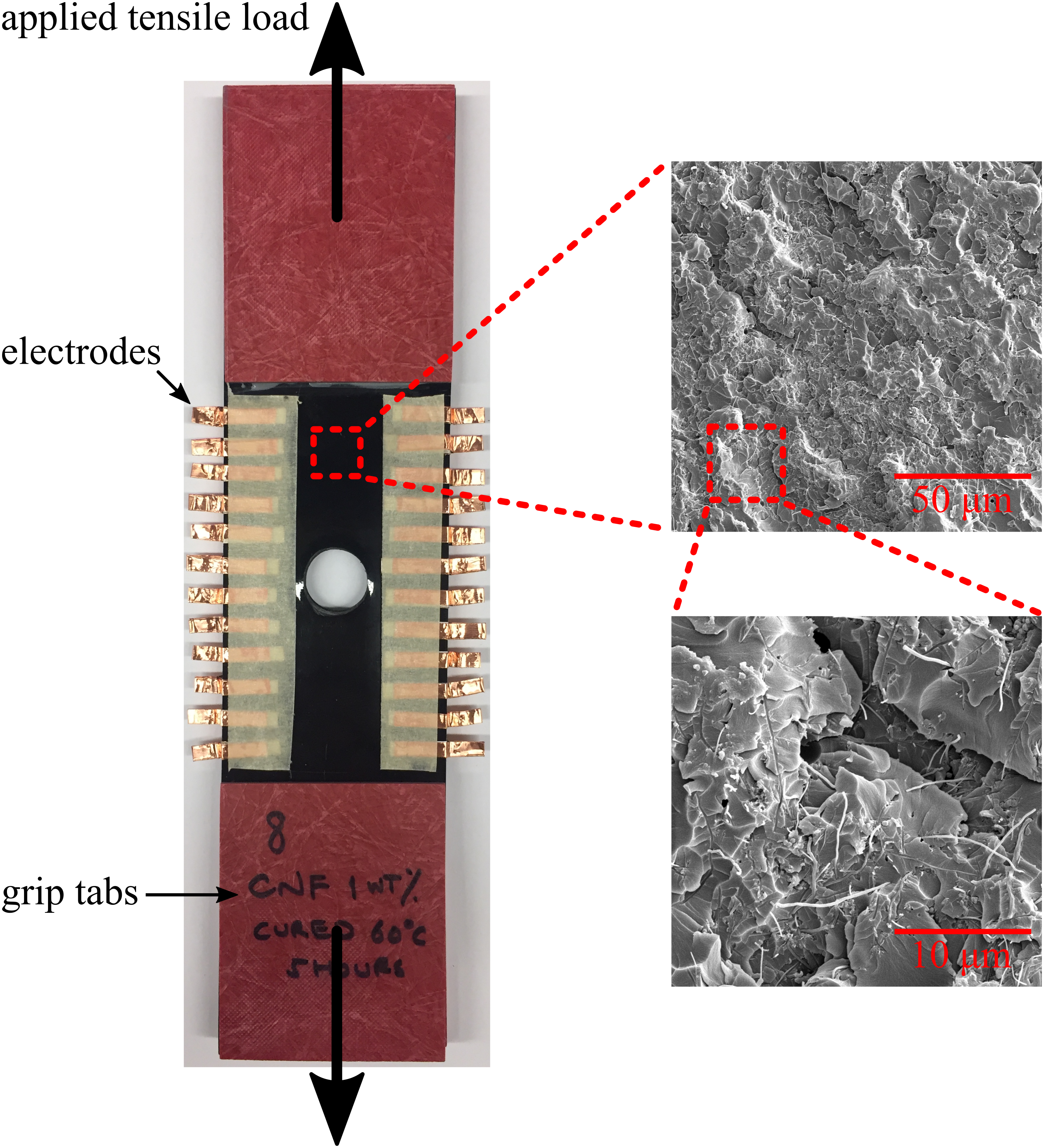}
\caption{Left: Experimental test specimen which is loaded in tension to a displacement of 0.75 mm. Right: SEM images of the underlying CNF nanofiller network.}
\label{experiment}
\end{figure}

In this subsection, we focus on the reconstruction of conductivity changes $\Delta \gamma$ in a piezoresistive nanocomposite subjected to external stresses. Specifically, a carbon nanofiber (CNF)-modified epoxy specimen was cast in the shape of a rectangular plate containing a stress raiser as shown in Fig. \ref{experiment} and loaded in a load frame under displacement-controlled settings up to a cross head displacement of 0.75mm. EIT was then used to image the conductivity change distribution of the plate. The nanocomposite was produced by dispersing CNFs at 1.0\% by weight in liquid epoxy resin using a combination of a chemical surfactant, a viscosity-reducing agent, high-energy centrifugal mixing, and bath sonication. Post-dispersion, the viscosity-reducing agent was evaporated out, air-release and curing agents were added, and the mixture was degassed at room temperature before being poured into a mold of the rectangular plate to cure. After manufacturing the CNF/epoxy plate, left over material was imaged via scanning electron microscope (SEM) to assess the CNF dispersion as also shown in Fig. \ref{experiment}.

EIT measurements were collected in both the undeformed configuration and during the application of the tensile load using an across injection scheme. That is, current was injected and grounded at electrodes directly across from each other (on the left and right edges) as voltages were likewise measured between across electrodes. This injection scheme was used to ensure that current interacted with conductivity changes near the hole at the center of the plate. Interested readers are directed to reference \cite{hassan2019failure} for a complete description of the material manufacturing and experimental EIT methods. Under these loading conditions, a positive tensile stress/strain concentration is expected to the left and right of the hole whereas a compressive concentration above and below it. For a material with positive piezoresistivity, this state of stress/strain is expected to elicit a conductivity decrease to the left and right of the hole and a conductivity increase above and below it in the piezoresistive CNF/epoxy.

To test the feasibility of the NN modelling error approximation approach in the context of EIT, we compare the reconstruction frameworks described in Eqs. \ref{NLDIm} and \ref{NLDIae} employing linear triangular finite elements for the reduced-order forward model $\overline{U}(\bar{\gamma})$.
For the higher-order (accurate) FEM model, we use quadratic triangular finite elements.
In our application, we use the higher-order model to (a) generate $\mathcal{A}$ and (b) as an accurate forward model in Eq. \ref{NLDIm} to be used as a baseline for comparing the reduced-order and error-compensated reconstructions.

\begin{figure}
\centering
\hspace{-8.5cm}
\includegraphics[width=8.5cm]{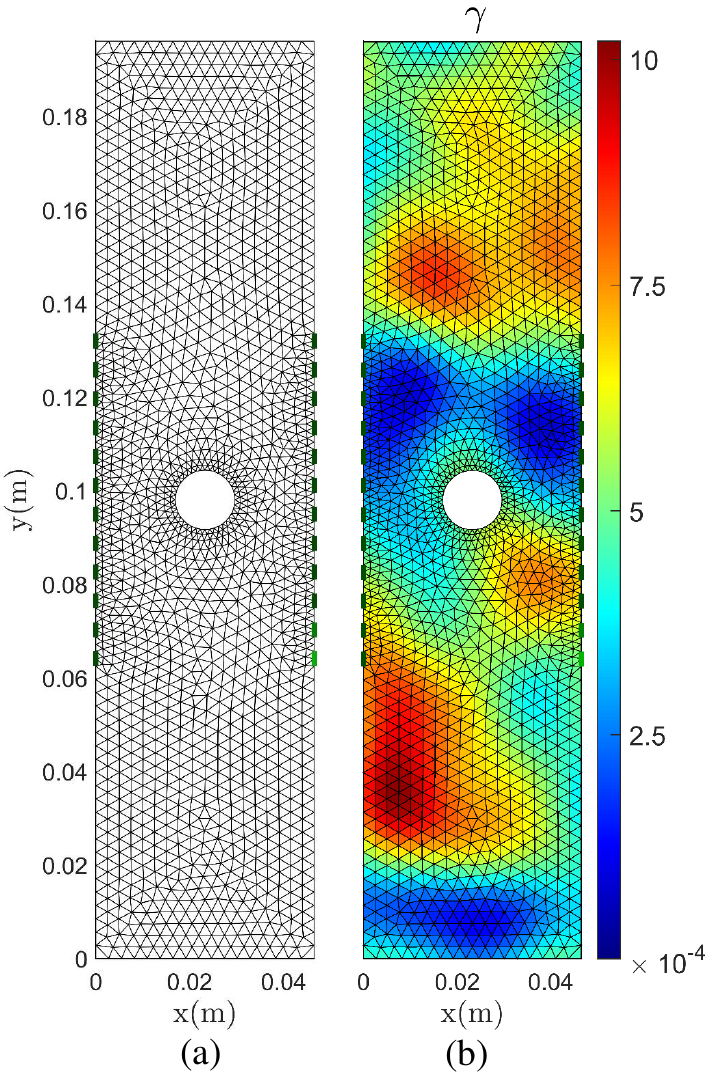}
\caption{EIT FE and data generation parameters, (a) triangular discretization consisting of 3050 elements used in the reduced- and high-order forward models and (b) an example of a random conductivity  distribution $\gamma$ (S/m) used in generating training data, i.e. a deformed absolute conductivity distribution. Boundary electrodes are shown in green.}
\label{mesh}
\end{figure}

In the numerical approach, we utilize the same triangularization, shown in Fig. \ref{mesh}(a), for all finite elements models.
The mesh consists of 3050 elements with a maximum element dimension of 3mm.
This discretization selection implies that all modelling errors (discrepancies between the reduced- and high-order forward models) result from the order of the interpolating polynomial (i.e. $p-$refinement \cite{surana2016f}).
To generate the model-error training data, we use the same number of training and validation data described in section \ref{NMDEl}.
To acquire the data, EIT simulations were generated from random blobby conductivity distributions (cf. Fig. \ref{mesh}(b) showing a deformed absolute conductivity distribution) using the same measurement and current injection protocol used in the experiments.
The range of the training data conductivity distributions was prescribed using a piezoresistive FEM simulation computing the expected conductivity change in a homogeneous specimen with conductivity $\gamma_\mathrm{ref}$ based on \cite{tallman2013arbitrary}. The expected conductivity change was then added to $\gamma_\mathrm{ref}$ in generating conservative bounds on $\gamma$ for the training samples: $10^{-4} \leq \gamma \leq 10^{-3}$ (S/m).
Lastly, the neural network architecture used for $\mathcal{A}$ mirrors the network described in section \ref{NMDEl}; however, the neurons used per hidden layer are reduced to 150 to avoid over fitting and reflect the smaller input and output sizes (132 inputs and outputs).

In solving the reduced-order forward model, high-order forward model, and NN modelling error compensated inverse problems the following numerical approaches are taken.
Firstly, the calculation of the Jacobian is conducted using second-order central differencing where the derivatives of $\mathcal{A}$ and $\overline{U}(\bar{\gamma})$ are calculated in tandem using parallel computations.
Importantly, it should be noted that the computational overhead in computing $\mathcal{A}$ and its derivative is minimal (herein, adding approximately 1-2\% to the total computing time per iteration).
Secondly, the constraints are handled using the interior point method and cubic barrier functions.
Lastly, iterations are continued until the change in the cost functionals between successive iterations is less than $10^{-4}$.

\begin{figure}[h!]
\centering
\hspace{-16cm}
\includegraphics[width=16cm]{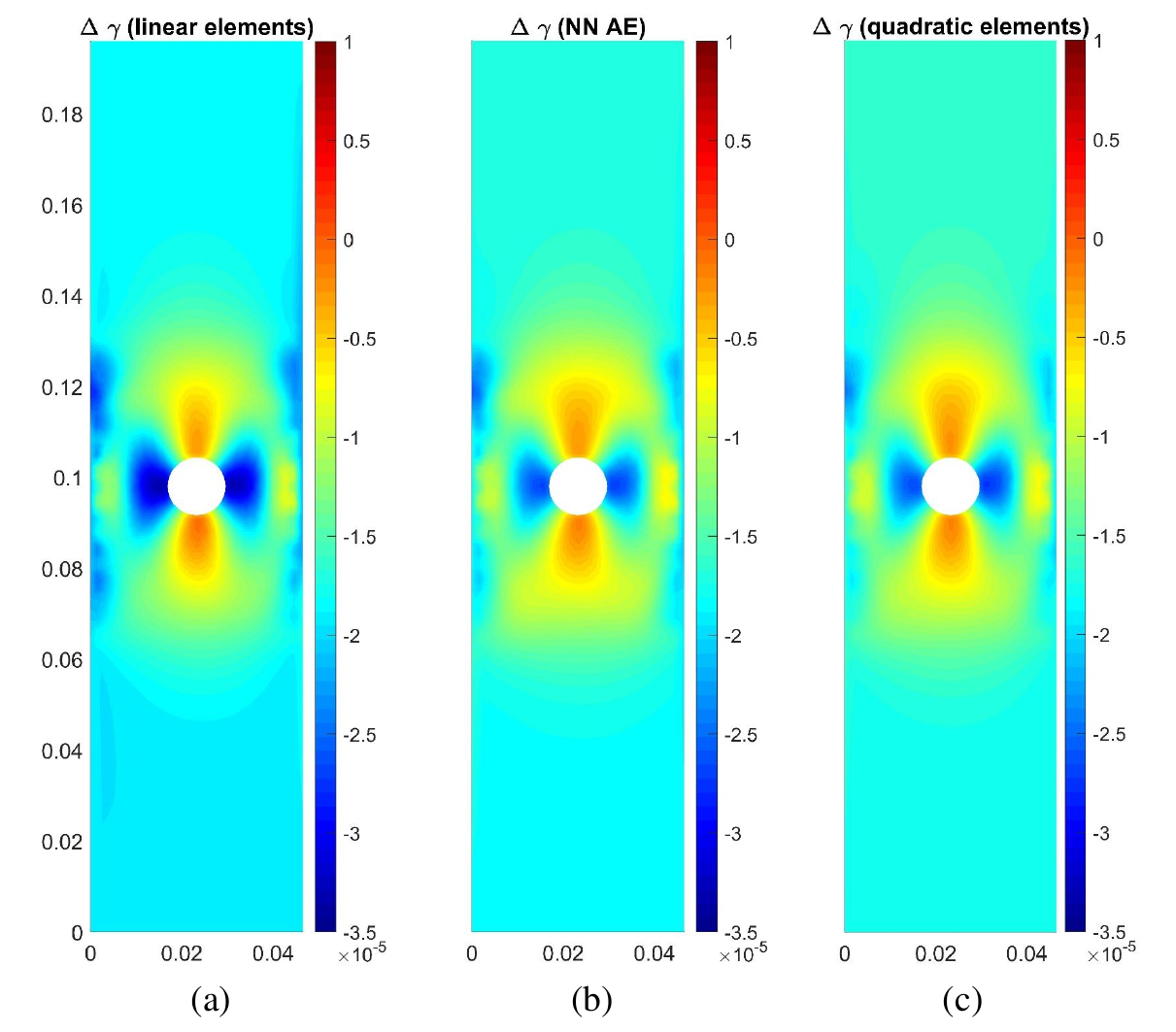}
\caption{EIT reconstructions reporting the change in electrical conductivity $\Delta \gamma \mathrm{(S/m)}$ for a nanocomposite material using (a) a reduced-order FE forward model with linear elements, (b) a NN approximate error (AE) compensated forward model, and (c) an increased-accuracy forward model with quadratic elements.
The units of length are meters.}
\label{piezores}
\end{figure}

EIT reconstructions reporting $\Delta\gamma$ from the piezoresistive voltage measurements are shown in Fig. \ref{piezores}.
Upon comparing images in Figs. \ref{piezores}(a) and (b) to the baseline higher-order forward model reconstruction in Fig. \ref{piezores}(c), it is immediately visually apparent that the NN modelling-error compensated reconstruction has fewer edge artifacts and more closely matches the baseline reconstruction.
In addition to this, it is clear that the background of the NN error compensated reconstruction much more closely matches the higher-order reconstruction.

To more quantitatively compare the reconstructions, we compute the MSE of images in Figs. \ref{piezores}(a) and (b) relative to the baseline image.
We find the MSEs to be $4.82 \times 10^{-6} \mathrm{(S/m)}$ and $6.21 \times 10^{-6} \mathrm{(S/m)}$ for the NN error compensated and reduced-order reconstructions, respectively.
In other words, the use of NN error compensation resulted in approximately 22\% reduction in MSE.
This result therefore supports the feasibility of using NN modelling error compensation in cases where reduced-order EIT forward models are used in reconstruction.
Further, this result indicates that NNs have potential to be more broadly applied for modelling error compensation in solving nonlinear ill-posed inverse problems in experimental settings.

\begin{figure}[h!]
\centering
\hspace{-15.2cm}
\includegraphics[width=15.2cm]{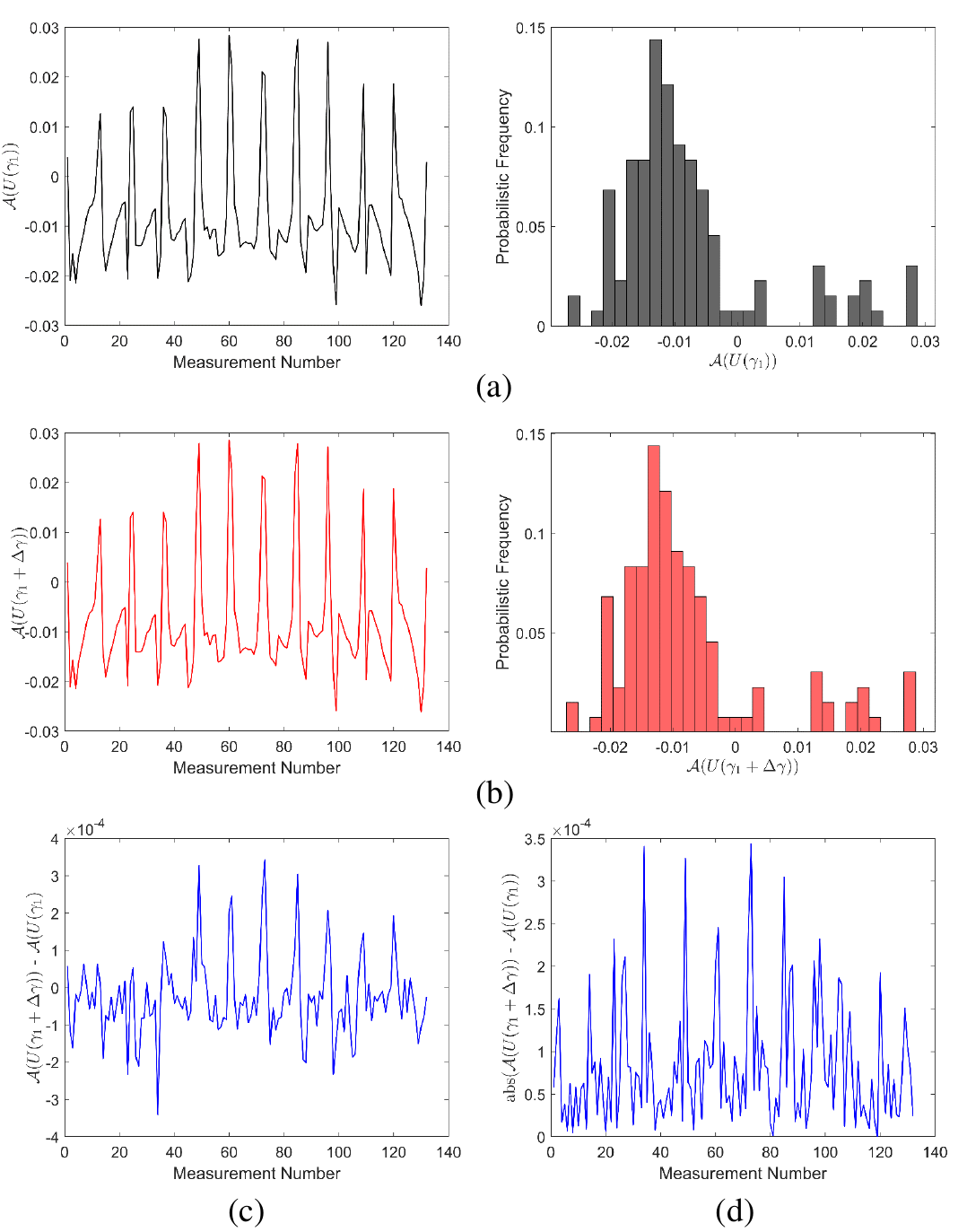}
\caption{
Neural Network error predictions for the nanocomposite imaging example, including (a) $\mathcal{A}(\gamma_1)$, corresponding to the unstressed conductivity state; (b) $\mathcal{A}(\gamma_1 + \Delta \gamma)$, corresponding to the stressed conductivity state; (c) difference between state error predictions $\mathcal{A}(\gamma_1) - \mathcal{A}(\gamma_1 + \Delta \gamma)$; and (d) absolute value difference between state error predictions abs($\mathcal{A}(\gamma_1) - \mathcal{A}(\gamma_1 + \Delta \gamma)$).
}
\label{errorsexp}
\end{figure}

In particular, it should be highlighted that the NN modelling error approach is intended to capture errors having potentially non-Gaussian distributions.
In the present example, we find that the errors captured by the NN are also non-Gaussian -- as reported in Figs. \ref{errorsexp}(a) and (b), reporting error prediction line plots and histograms for $\mathcal{A}(\gamma_1)$ and $\mathcal{A}(\gamma_1 + \Delta \gamma)$, respectively.
Indeed, by examining the histogram in Figs. \ref{errorsexp}(a) and (b), we observe NN predictions having errors that would be poorly approximated by a conventional uni-modal Gaussian distribution.
Further, and more subtly, we note that the differences between error predictions for each state ($\gamma_1$ and $\gamma_2$) are small, as evidenced by Figs. \ref{errorsexp}(c) and (d).
This observation relays the importance of the NN modelling error prediction accuracy.
In the context of an ill-posed problem, such as EIT, even small changes in the error predictions may have large effects on the recovered conductivity.

Notwithstanding the initial successes reported in this subsection, it is important to note the limitations of the proposed EIT NN model error estimator.
Firstly, the NN was trained using data specific to the problem solved, i.e. using prior information on the expected range of $\gamma$ and knowledge of the exact geometry/measurement protocol.
This means that the NN is not generalizable to accurately estimate modelling errors for EIT problems with different dimensions, measurement patterns, or significantly varying ranges of $\gamma$.
Rather, the NN used is \emph{specialized} to the EIT problem at hand and is therefore the primary disadvantage.
On the other hand, when an EIT geometry and stimulation protocol are fixed, it may be possible to feasibly extend the range of the conductivity distributions used during training to develop a more general NN modelling error compensation model.
Lastly, the use of projection models coupled with the proposed NN error compensation approach may improve performance and generality of the NN for use in error-compensated EIT imaging.
Furthermore, in future work, we aim to investigate (i) the latter two models and (ii) the use of the proposed method in cases where boundary uncertainty is present.

\subsection{ {Robustness of modelling error predictions to network parameters}}
 {In the previous subsection, the efficacy of implementing NN model error predictions in an EIT reconstruction framework was demonstrated.
It remains to question, however, the extent to which important neural network model parameters effect error predictions.
Such sensitivities may have significant implications on the generalizabilty of NN model error estimators and, consequentially, are investigated in this subsection to better understand the effects of network parameters on predicted errors.
To do this, we vary three key network parameters and analyze their effects on modelling error metrics (cf. section \ref{NMDEl} for metric details), specifically: (a) the number of hidden layers, (b) the number of neurons per hidden layer, and (c) the activation function type.
}

 {In this study, we simulate 6000 random conductivity distributions using the same geometry, accurate/reduced FE discretizations, EIT simulation/measurement pattern, and sampling protocol described in the previous subsection.
From the 6000 samples, 5000 are used for training and the remaining 1000 are used for testing error predictions.
Moreover, we train different neural networks having 50, 100, 200, or 500 neurons and vary the number of hidden layers between one and three.
For each of these networks either ReLU or LeakyReLU (negative region slope of 0.1) activation functions are used.
As a result, a total of 24 different networks are tested using the same training and testing data.
}

\begin{sidewaysfigure}
  \centering
  \hspace{-24cm}
  \includegraphics[width=26cm]{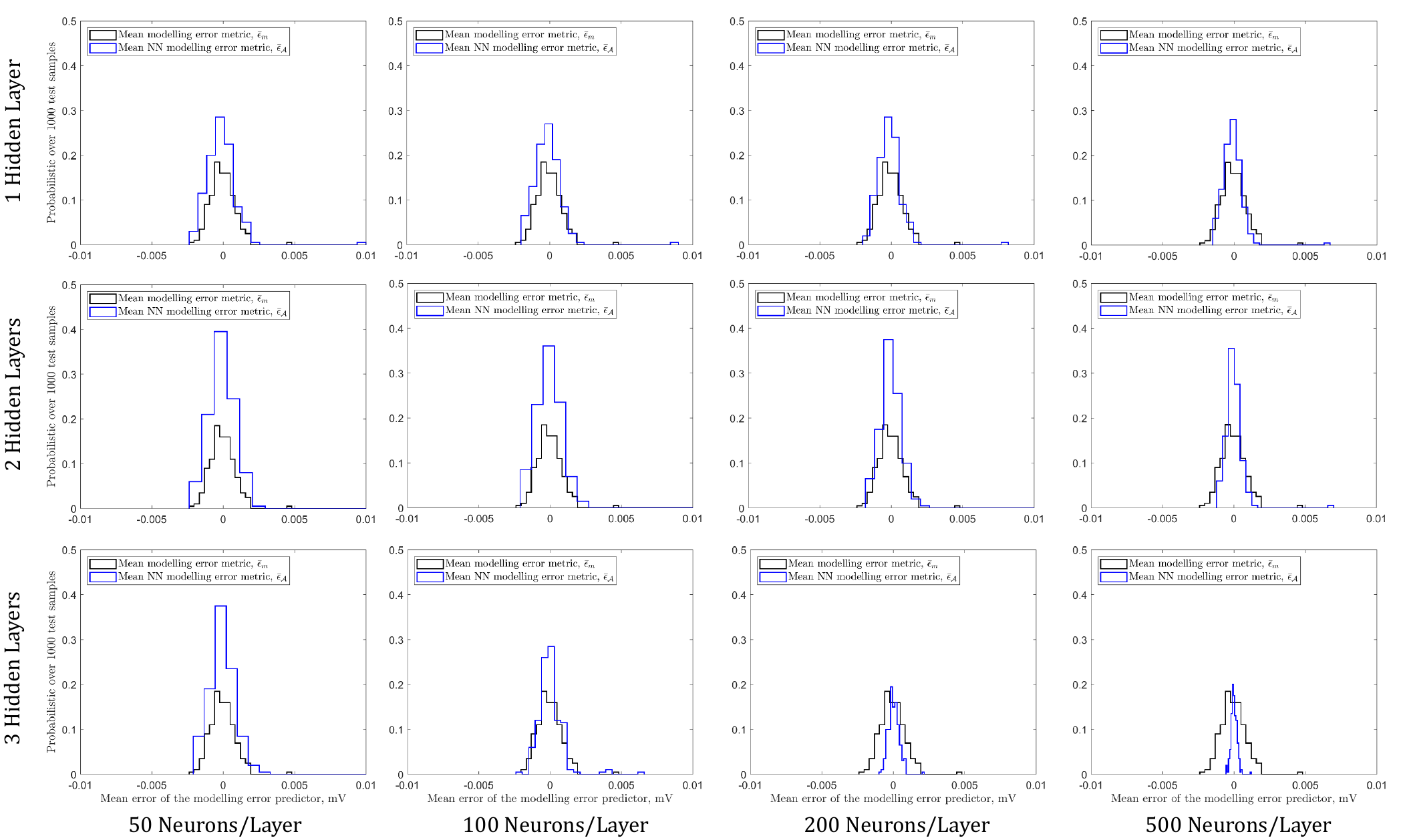}
  \caption{ {EIT forward model error prediction metrics for different ReLU activated NN architectures compared against mean model error prediction metrics.
  Histograms with blue edge coloring represent NN model error prediction metrics, while black edge coloring represents mean model error metrics.}}
  \label{ReLUFig}
 
\end{sidewaysfigure}

 {We begin our investigation by analyzing the NN model error prediction metrics resulting from architectures utilizing ReLU activation functions.
These metrics are compared with the mean model error predictions, as shown in Fig. \ref{ReLUFig}.
We first note that in order for a given NN modelling error predictor to outperform a mean modelling error predictor, we would expect that the NN error prediction metric is more closely grouped about zero than the mean modelling error predictor.
This is only obvious in two cases where the number of hidden layers (three) and number of neurons per layer are largest (200 and 500).
This indicates that ReLU activated NN error predictors do not clearly outperform the mean modelling error predictor when the networks are shallow and have few neurons (specifically, when the number of neurons per layer is less than the output size).
However, in the case where both the network depth and number of neurons are largest, the standard deviations of the NN and mean modelling error metrics are 2.7$\times 10^{-4}$mV and 8.7$\times 10^{-4}$mV, respectively.
This result shows that, when the ReLU activated NN error predictor architecture is chosen properly, the deviation from zero error in NN modelling error predictions is significantly lower than the standard mean modelling error predictors.
}

\begin{sidewaysfigure}
  \centering
  \hspace{-24cm}
  \includegraphics[width=26cm]{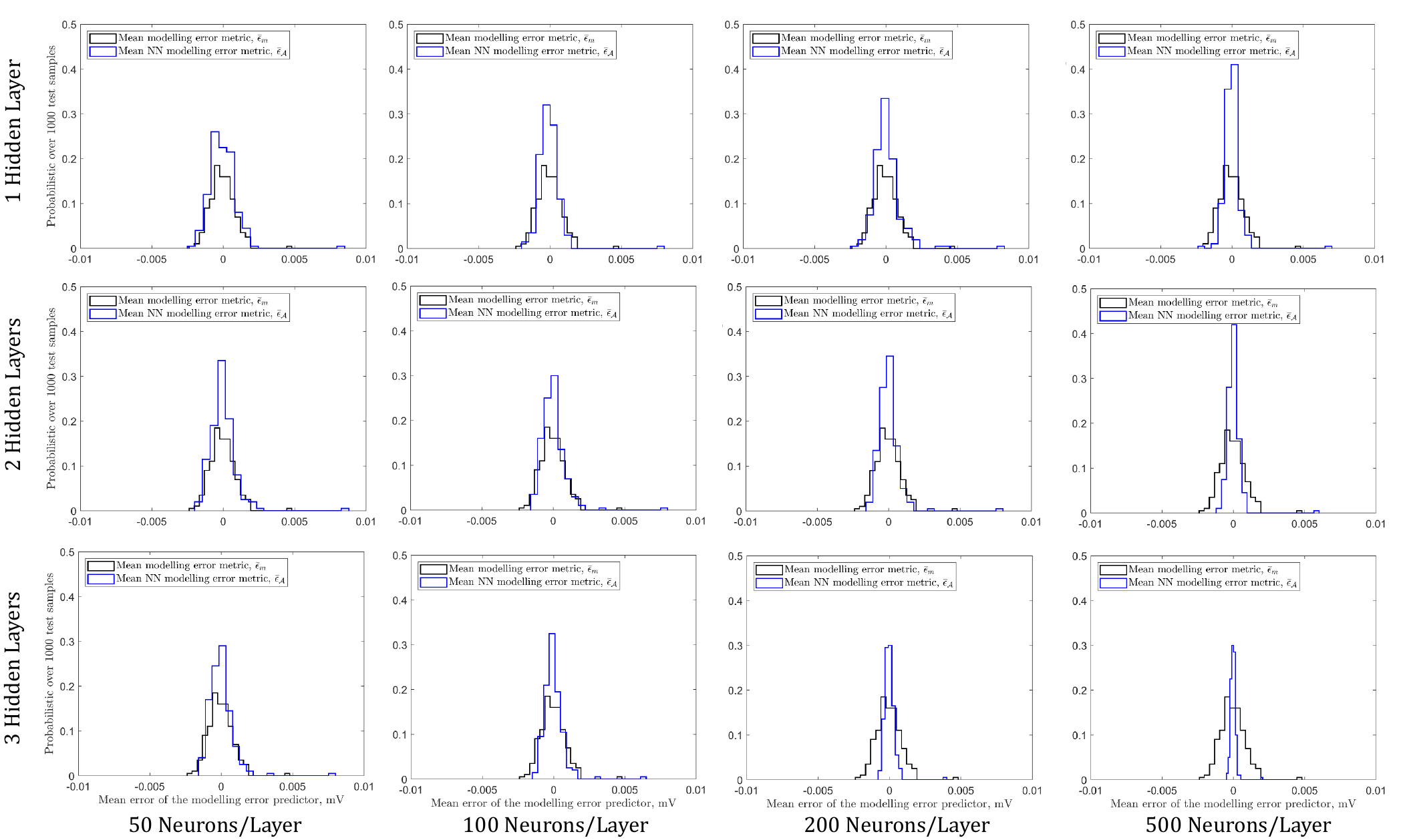}
  \caption{ {EIT forward model error prediction metrics for different LeakyReLU activated NN architectures compared against mean model error prediction metrics.
  Histograms with blue edge coloring represent NN model error prediction metrics, while black edge coloring represents mean model error metrics.}}
  \label{LeakyReLUFig}
 
\end{sidewaysfigure}

 {In comparison with the previously reported ReLU modelling error metrics, the LeakyReLU activated modelling error metrics across most network architectures studied herein compare favorably as shown in Fig. \ref{LeakyReLUFig}.
In general, we observe that fewer neurons and hidden layers are required to outperform the mean modelling error predictor.
The improvements in LeakyReLU activated NN model error predictors are most likely due to the fact that the use of LeakyReLU functions permits small non-zero gradients, while the use of ReLU functions can result in saturation (zero gradients) \cite{maas2013rectifier}.
Nonetheless, we believe this finding holds promise for future NN error prediction applications, whereby NN architectures can be specifically tailored to compensate/quantify modelling errors.
}

\begin{figure}[h!]
\centering
\hspace{-12cm}
\includegraphics[width=12cm]{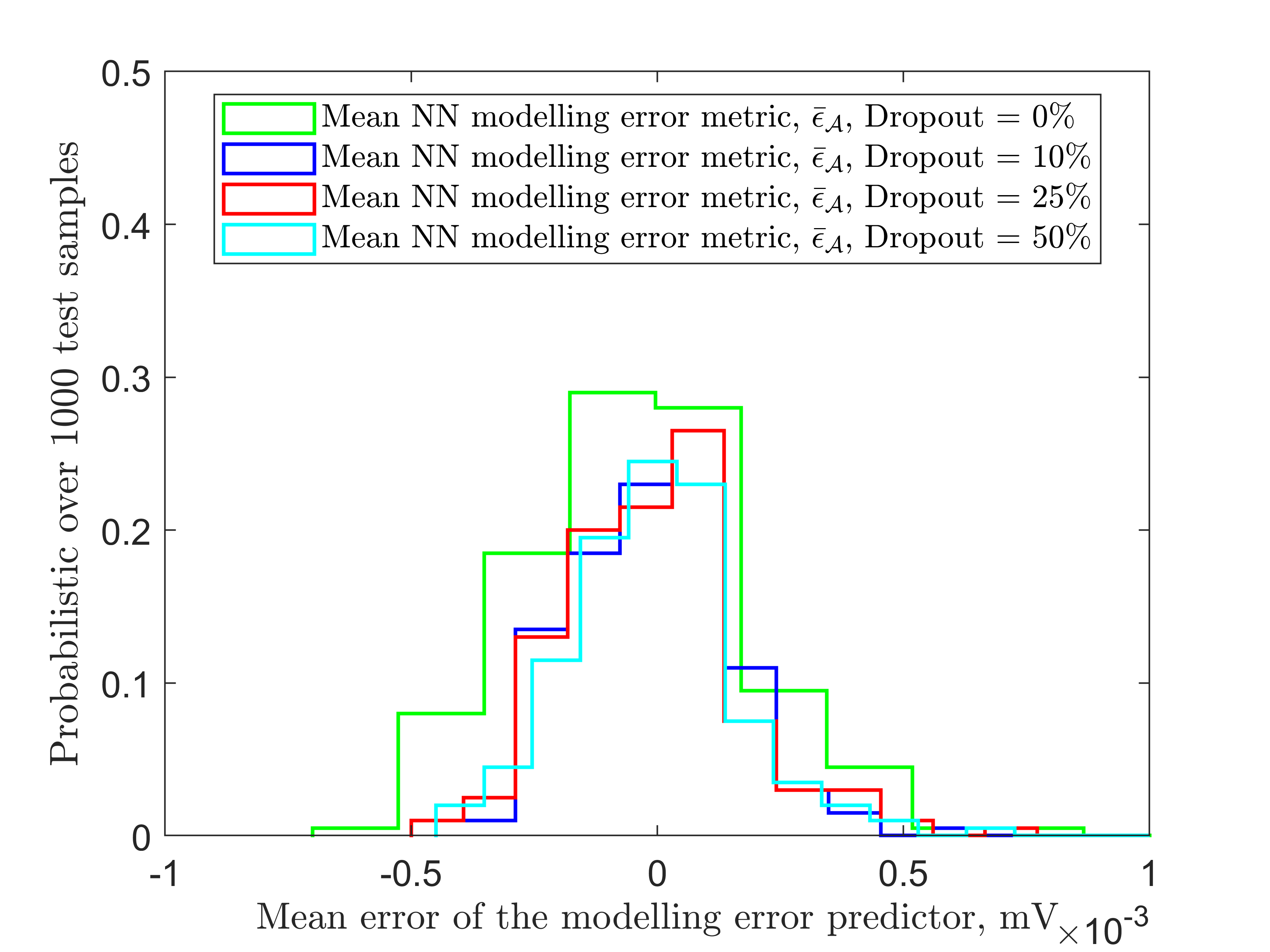}
\caption{ {The effect of dropout rate on model error prediction metrics for a NN with three hidden layers, 500 neurons per layer, and LeakyReLU activation functions.}
}
\label{dropout}
\end{figure}

 {Lastly, it is worth noting that the networks' ability to accurately predict modelling errors is significantly influenced by regularization.
In our network training to this point, the combined use of dropout and adaptive computing of the regularizing hyperparameter $\lambda$ (Eq. \ref{NNr}) was adopted. 
Using the best performing NN for error prediction in this study (three hidden layers and 500 neurons per layer), we studied the effect of dropout rate on model error predictions using the training and testing data described in this subsection.
The results from this dropout sensitivity analysis are shown in Fig. \ref{dropout}.
The major difference of note is the deviation between the 0\% dropout error metric and the 10-50\% dropout metrics -- where the 0\% dropout error metric has a significantly larger standard deviation.
On the other hand, error metrics for dropout rates between 10\% and 50\% differ by a maximum of 8\% standard deviation indicating little performance differences between these NNs.
As a whole, we can conclude that error prediction networks trained without dropout herein are (a) significantly less generalizable than those incorporating at least 10\% dropout and (b) under-regularized.
Meanwhile, the marginal error prediction performance differences for networks trained with 10-50\% dropout are likely due to the effectiveness of the adaptive Levenberg-Marquardt regularization framework.
}

\break

\section{Discussion and conclusion}
The aim of this work was to propose a straightforward neural network-based approach capable of accurately approximating and compensating for numerical modelling errors in augmented direct and inverse problems.
A strong motivator behind this effort was the general need for model error compensation in the ever-present situation where modelling errors are non-Gaussian.
To these aims and motivations, we affirmatively demonstrated the viability of using neural networks for both approximating such modelling errors in direct problems and compensating errors in ill-posed inverse problems.

A central result of this study showed that the neural network model error approximator was effective in predicting modelling errors for static PDEs discretized using the FEM.
One inherent limitation of the neural network approach, however, was that the modelling error predictions made by the neural network are only valid if the PDE solution input into the neural network does not exceed the training data.
In cases where the input data does exceed the space of the training data, the proposed approach will possibly not provide accurate estimates of modelling errors.
In this sense, it was noted that the data provided to train the neural network is, in fact, prior information informing the realm of possible error predictions. 

This subjectivity has advantages and disadvantages.
On one hand, if the numerical solutions of a PDE are known to lie within a very tight bound based on, e.g. physical realizations, then the neural network may be very suitable for a broad suite of that PDE's solutions.
On the other hand, if the solutions of a PDE vary by many orders of magnitude, the space of the training data must be very large and the resulting error predictions may be comparatively less accurate than the former case.
In either case, the proposed approach to predicting modelling errors is highly \emph{specialized}, in that, our formulation was only applicable to a specific geometry and boundary conditions -- which would certainly need to be considered in cases where, for example, problem conditions change continuously.

Further to these realizations, neural network model error compensation was shown to be useful in a highly nonlinear ill-posed inverse problem (EIT) while contributing marginally to the computational overhead.
In the experimental program, we showed that an EIT reconstruction using reduced-order forward models with neural network error compensation closely matched reconstruction employing higher-order FE forward models.
While this result does demonstrate that reconstructions using neural network error compensation \emph{can} improve solutions to inverse problems, we do not insinuate that this is true in all cases.
The fundamental basis the former lies in the fact that, when forward operators are of sufficiently high accuracy (including error compensated models), inverse solutions may in fact degrade -- as shown in \cite{smyl2019less}.
In essence, this reality is well encapsulated in quote from \cite{burger2019convergence} stating, ``errors in the data should not converge faster than the errors in the operator ... there is no need to measure something more precisely than we can predict it."

From an inverse problems standpoint, this raises an interesting question: ``can [neural network] error compensation be so effective that solutions are consistently degraded?"  
While this is an intuitively interesting query, we are far from reaching this point.
Much work is still needed to improve the generality of neural network error approximators in order for them to be broadly applicable enough for this question to be meaningful in the wider sense -- e.g., addressing the noted issues related to the proportional accuracy of NN error predictions with respect to the magnitude of reduced order solutions.
In closing, however, the results presented herein are promising from a pragmatic standpoint, affording potential opportunities for scientists and engineers to, e.g., (i) compensate/augment reduced-order models when computational speed-ups are required, (ii) rapidly predict model error statistics, (iii) improve solutions to \emph{specialized} inverse problems at little additional computational overhead, and (iv) provide insights into non-Gaussian model errors and their causality.

\section*{Acknowledgments}
DL was supported by National Natural Science Foundation of China (Grant No. 61871356).
This work was partially supported by the Academy of Finland Project 312123 (Finnish Centre of Excellence in Inverse Modelling and Imaging, 2018--2025).

\bibliography{bibliography}

\end{document}